\documentclass[aip,apl,amsmath,amssymb,reprint]{revtex4-2}

\usepackage{graphicx}
\usepackage{dcolumn}
\usepackage{bm}
\usepackage{booktabs}
\usepackage{microtype}
\usepackage{hyperref}
\usepackage{enumitem}
\usepackage{color}

\newcommand{\mpc}[1]{{\color{black}{{#1}}}}

\begin{document}

\title{The Glass Transition: From Competing Theories to Discriminating Tests}

\author{Massimo Pica Ciamarra}
\email{massimo@ntu.edu.sg}
\affiliation{Division of Physics and Applied Physics, School of Physical and Mathematical Sciences, Nanyang Technological University, Singapore 637371, Singapore}

\date{\today}

\begin{abstract}
The dramatic slowing down of liquids approaching the glass transition remains unresolved, partly because competing theoretical explanations are difficult to discriminate. Indeed, several mature, contrasting frameworks account for the main features of glassy phenomenology, making agreement with familiar observations a weak basis for identifying the microscopic mechanism responsible for the slowdown.
Recent numerical advances are changing this situation in two complementary ways. Deep equilibration extends controlled equilibrium studies into regimes previously accessible only by extrapolation, allowing established predictions to be stress-tested over a broader dynamical range. At the same time, new numerical methods increasingly provide direct access to theory-specific quantities such as configurational entropy, static lengths, elastic properties, and local activation barriers. I discuss how these   developments constrain thermodynamic, facilitation-based, elastic, and excitation-based descriptions. The emerging picture is not yet a unique theory, but a set of increasingly discriminating tests that can reveal where apparently successful laws cease to be universal and identify the microscopic ingredients that causally control the activation barrier.
\end{abstract}

\maketitle

\section{The problem has changed}

As a liquid approaches the glass transition upon cooling, its structural relaxation time may increase by many orders of magnitude while conventional measures of structure change only modestly. The central theoretical question is why this apparently limited structural evolution accompanies such a dramatic dynamical slowdown.

Historically, mode-coupling theory (MCT) provided a microscopic account of the initial rapid slowdown in terms of nonlinear feedback among density fluctuations. 
\mpc{In its ideal form, MCT predicts that the structural relaxation time diverges at a finite temperature $T_{\rm MCT}$, signaling a dynamical arrest. 
In finite-dimensional liquids, however, this ideal transition is avoided and relaxation remains possible below $T_{\rm MCT}$, notably through activated processes~\cite{jan18,ber11}.}
Generalized MCT seeks to address limitations of standard MCT by systematically incorporating higher-order density correlations~\cite{janssen16gmct,luo22gmct}. The temperature $T_{\rm MCT}$ is commonly used as a reference point for the crossover to the deeply supercooled regime, $T<T_{\rm MCT}$, in which activated relaxation becomes increasingly important.

Several frameworks address this activated regime. Building on the Adam--Gibbs relation~\cite{ada65}, random first-order transition (RFOT) theory attributes the slowdown to a competition between configurational entropy and the mismatch cost between amorphous states, leading to a growing static length~\cite{kir89,lub07,bir23}. Dynamic facilitation instead places localized mobility excitations and the constraints on their propagation at the center of the description, without requiring a thermodynamic transition~\cite{rit03,cha10,keys2011excitations}. Elastic models relate the growing barrier to the increasing mechanical cost of a localized rearrangement, from the shoving model to microscopic cage-plus-elastic formulations such as the Elastically Collective Nonlinear Langevin Equation (ECNLE)~\cite{dyr96,dyr06,mir14a,mir14b,pha18}. A complementary excitation-based viewpoint focuses directly on the spectrum and geometry of localized barriers and on the interactions among such events~\cite{cia24,jic25}. For a recent review of elastic and local-excitation approaches, see Ref.~\onlinecite{PicaCiamarra2026local}.

The field is therefore no longer short of plausible mechanisms. The difficulty is almost the opposite: mature theories reproduce overlapping sets of signatures. Dynamical heterogeneity, non-Arrhenius relaxation, decoupling between transport coefficients, stretched correlation functions, and growing dynamical length scales are not unique fingerprints of one framework~\cite{ber11,biroli23dh}. Indeed, descriptions based on configurational entropy, collective strings, elastic shoving, localization, and parabolic laws can all account reasonably well for the same relaxation data~\cite{Yuan2024}. Fit quality and generic glassy phenomenology are consequently weak discriminators.

Distinguishing among these mechanisms requires tests that go beyond fitting relaxation data or identifying generic glassy signatures. 
Recent advances in simulation make such tests increasingly feasible. Continued increases in computational reach, together with swap Monte Carlo~\cite{grigera01,ninarello17} and other enhanced-sampling methods, now extend equilibrium studies into regimes that were previously inaccessible.
This opens two complementary routes to sharper tests. First, a successful law can be followed beyond the window in which it was established and tested where its extrapolation becomes a genuine prediction. Second, quantities identified by a theory as microscopic ingredients can increasingly be measured independently and confronted directly with the predicted barrier or relaxation time. The excitation-spectrum result discussed below provides a particularly direct example of the latter strategy, while the low-temperature fate of the parabolic law illustrates the former. This Perspective does not aim to review these frameworks---recent accounts are
available~\cite{ber11,bir23,lub07,cha10,dyr06,PicaCiamarra2026local}---but to use both routes to reassess what current successes establish, where quantitative tensions are emerging, and, for each framework, which independently measurable ingredients now permit quantitative, discriminating tests of the dynamics.

\begin{figure*}[!t]
\centering
  \includegraphics[width=\textwidth]{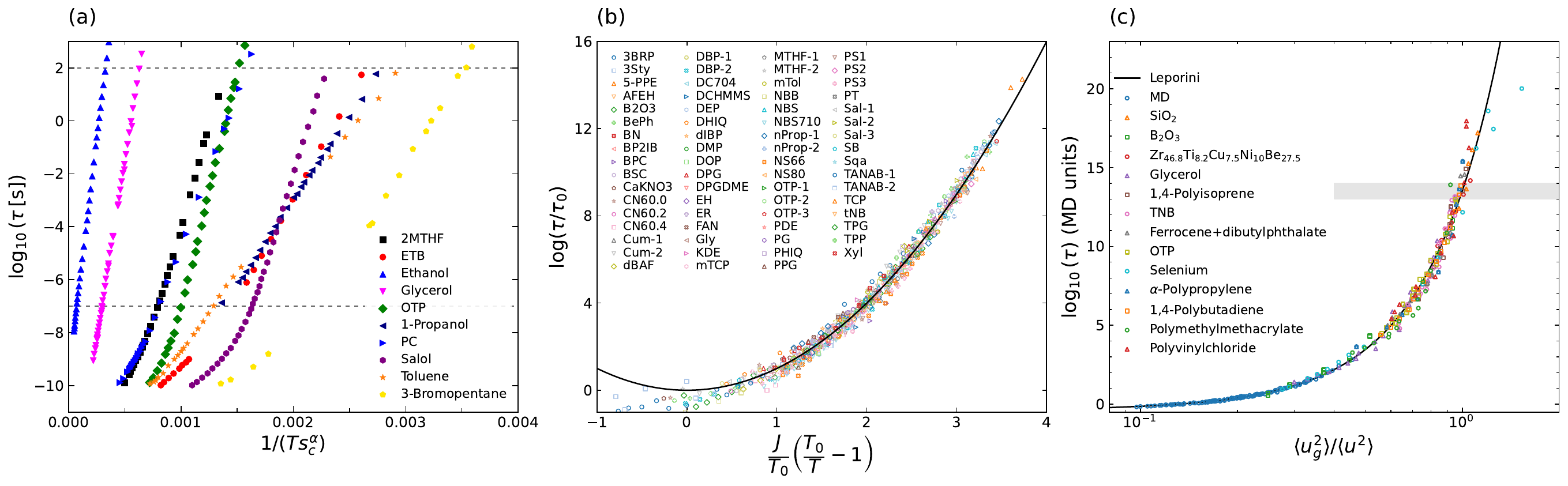}
  \caption{  
  Representative ``success plots'' for three theoretical approaches to the glass transition.
  (a) Generalized RFOT representation, $\log_{10}\tau$ versus $1/[T s_c^{\alpha}]$, for molecular glass formers.  The approximate linearity is consistent with generalized RFOT scaling; extrapolation to $1/(T s_c^{\alpha})=0$ fixes the effective microscopic intercept $\log_{10}t_0$. 
  Data from Ref.~\citenum{oza19}.
  (b) Scaled relaxation data for a large number of glass formers described by the parabolic law associated with hierarchical dynamic facilitation. 
  Data from Ref.~\citenum{elm09}.
  (c) Relaxation-time and viscosity data, vertically shifted onto the molecular-dynamics (MD) master-curve scale, versus the reduced Debye--Waller factor $\langle u_g^2\rangle/\langle u^2\rangle$;  the solid line is the prediction proposed there, and the gray band marks the glass-transition region in MD time units.
  Data from Ref.~\citenum{lar08}.
  Taken separately, each representation is compelling; taken together, they illustrate why agreement with $\tau(T)$ alone does not uniquely identify a microscopic mechanism.}
  \label{fig:success}
\end{figure*}

\section{What do theories actually disagree about?}
At sufficiently low temperature, structural relaxation is commonly represented in activated form,
\begin{equation}
\tau(T)=t_0\exp\left[\frac{E(T)}{T}\right],
\label{eq:tau}
\end{equation}
where $\tau(T)$ is the structural ($\alpha$) relaxation time, $t_0$ is a microscopic timescale, and $E(T)$ is an effective activation barrier.
\footnote{Here and throughout, temperature is expressed in energy units.}
\mpc{Operationally, $\tau$ characterizes the decay of structural correlations. In simulations and scattering experiments, for example, it can be defined from the decay of a density correlation function, such as the intermediate scattering function, to a fixed fraction of its initial value. In molecular liquids, $\tau$ also closely tracks the viscosity: within a Maxwell description, $\eta \simeq G_\infty \tau$, so that the relaxation rate $1/\tau$ is approximately proportional to the inverse viscosity.}
The dramatic super-Arrhenius growth of $\tau$ is then associated primarily with the increase of $E(T)$ upon cooling. The main theories disagree on the physical origin of this increase. For the discussion below, it is useful to organize this disagreement around two largely independent questions:
\begin{enumerate}
\item Is thermodynamic organization required for the growth of the activation barrier, or can the slowdown be generated without it?
\item Are growing correlations part of the mechanism that raises the barrier, or do they emerge because increasingly slow local rearrangements interact and facilitate one another?
\end{enumerate}
\mpc{
These distinctions should not be taken to imply that the different ingredients are necessarily mutually exclusive. Connections between thermodynamic organization and collective dynamics have been explored, for instance, by relating configurational entropy to string-like cooperative motion~\cite{starr2013relationship,xu21douglas}, while possible connections between thermodynamic and facilitation-based descriptions have also been discussed~\cite{roy20,bir23}. Likewise, elastic interactions between localized rearrangements can themselves generate facilitation and dynamical heterogeneity~\cite{oza23}. The sharper question is therefore which of these ingredients controls the growth of the activation barrier, which are correlated consequences of the slowdown, and whether their relative importance may differ among glass formers or dynamical regimes.
}

\subsection{RFOT -- a growing thermodynamic length}
A major theoretical strength of RFOT is that its mean-field construction can be solved exactly for hard spheres in the infinite-dimensional limit, where metastable amorphous states, configurational complexity, and the dynamical and Kauzmann transitions are sharply defined~\cite{charbonneau14mf,parisi20}. How this mean-field
structure is modified in finite dimensions, and whether the resulting mosaic
construction is quantitatively valid, remain central open questions.

In finite dimensions, RFOT relates the super-Arrhenius slowdown to the
emergence of a growing static amorphous length.
Consider a region of linear size $\ell$. Rearranging this region into a different amorphous state involves a competition between the mismatch free-energy cost with its surroundings and the configurational-entropy gain associated with the multiplicity of alternative amorphous states~\cite{kir89,lub07,bir23}. At the scaling level, in dimension $d$,
\begin{equation}
\Delta F(\ell,T)
\sim
\Upsilon(T)\ell^\theta
-
T s_{\rm c}(T)\ell^d,
\label{eq:rfot_balance}
\end{equation}
where $s_{\rm c}(T)$ is the configurational entropy density, $\Upsilon(T)$ is the amorphous mismatch scale, and $\theta$ characterizes the scaling of the mismatch cost.

Balancing the two contributions defines the mosaic length, operationally associated with the point-to-set length $\xi_{\rm PTS}$. 
\mpc{The latter quantifies how far an amorphous boundary constrains the equilibrium structure: particles outside a cavity are frozen, the interior is re-equilibrated, and the overlap with the reference configuration is measured near the cavity center. The point-to-set length corresponds to the cavity size beyond which the influence of the frozen boundary is lost~\cite{BerthierCharbonneauYaida2016},}
\begin{equation}
\xi_{\rm PTS}(T)
\sim
\left[
\frac{\Upsilon(T)}
{T s_{\rm c}(T)}
\right]^{1/(d-\theta)}.
\label{eq:rfot_xi}
\end{equation}
If the temperature dependence of $\Upsilon(T)/T$ is weaker than that of $s_{\rm c}(T)$, the decrease of configurational entropy controls most of the growth of $\xi_{\rm PTS}$.

RFOT further assumes that structural relaxation proceeds through activated rearrangements correlated over this growing length,
\begin{equation}
E(T)
\sim
\Delta_0\,\xi_{\rm PTS}(T)^\psi,
\label{eq:rfot_barrier}
\end{equation}
where $\psi$ is a barrier exponent and $\Delta_0$ denotes a microscopic barrier scale, taken in the simplest scaling form to have a weaker temperature dependence than the growing collective factor. Combining Eqs.~\eqref{eq:rfot_xi} and \eqref{eq:rfot_barrier} gives
\begin{equation}
\ln\frac{\tau(T)}{t_0}
=
\frac{A}
{T\,s_{\rm c}(T)^\alpha},
\qquad
\alpha=\frac{\psi}{d-\theta},
\label{eq:rfot_ag}
\end{equation}
where $A$ is treated as temperature independent.

The original RFOT construction takes $\psi=\theta=d/2$, yielding $\alpha=1$ and the Adam--Gibbs form. More generally, RFOT allows a nontrivial exponent $\alpha$~\cite{oza19}. Ref.~\citenum{oza19} tested this generalized relation using configurational-entropy and relaxation-time data for molecular glass formers over the experimentally relevant dynamical regime. The conventional $\alpha=1$ form fails for most systems over the full range, whereas fitted values of $\alpha$ typically below unity produce the approximately linear plots shown in Fig.~\ref{fig:success}(a). This is a substantial success, but the fitted exponent and the associated intercept also provide quantitative constraints, to which we return below.

\subsection{Dynamic facilitation -- a growing length without thermodynamics}
\mpc{Dynamic facilitation has its roots in kinetically constrained models introduced by Fredrickson and Andersen, in which the equilibrium thermodynamics is essentially trivial while local rearrangements are allowed only in the presence of nearby mobility~\cite{fredrickson84,fredrickson85}. 
Related hierarchical constrained models were introduced by Palmer \emph{et al.} and, later, by J\"ackle and Eisinger~\cite{palmer84,jackle91}. }
These models demonstrated that dramatic slowing down and dynamical heterogeneity can arise from dynamical constraints alone, without invoking nontrivial thermodynamics~\cite{rit03,cha10}. 
The later atomistic formulation of Keys \emph{et al.} identified localized mobility excitations in supercooled liquids and a specifically hierarchical regime in which facilitation becomes increasingly directional on cooling~\cite{keys2011excitations}.


In that picture, localized mobility excitations retain a microscopic size while their concentration $c_a(T)$ decreases. The typical distance between excitations therefore grows,
\begin{equation}
\ell_a(T)\sim c_a(T)^{-1/d_f},
\end{equation}
where $d_f$ characterizes the geometry of facilitated relaxation. Relaxing over the increasing distance $\ell_a$ requires increasingly extended hierarchical sequences of events, leading to the parabolic law
\begin{equation}
\ln\frac{\tau(T)}{\tau_{\rm o}}
=J^2\left(\frac{1}{T}-\frac{1}{T_{\rm o}}\right)^2,
\qquad T<T_{\rm o},
\label{eq:parabolic}
\end{equation}
where $T_{\rm o}$ is the onset temperature and $\tau_{\rm o}=\tau(T_{\rm o})$. The onset marks the crossover below which transient localization and hierarchical facilitation become apparent. Here the growing dynamical length is part of the mechanism generating the increasing barrier, without invoking a growing thermodynamic length.

The empirical support for this picture is unusually broad. Ref.~\citenum{elm09} showed that transport data for more than fifty fragile glass formers collapse onto the parabolic form, as illustrated in Fig.~\ref{fig:success}(b), and the later atomistic analysis connected its parameters to excitation dynamics~\cite{keys2011excitations}. The resulting claim is therefore stronger than that of a convenient fitting law: the parabola is meant to express a broadly applicable hierarchical facilitation mechanism.

\subsection{Elastic and excitation-based pictures}
Elastic descriptions trace the slowdown to the increasing mechanical cost of localized rearrangements. Their conceptual roots go back at least to Goldstein, who argued that below an onset temperature relaxation proceeds through activated transitions between nearby minima of the potential-energy landscape. He pictured these transitions as spatially localized rearrangements associated with long-ranged elastic strain fields, which provide a mechanism for coupling different rearranging regions~\cite{gol69}. 

The simplest modern realization is Dyre's shoving model~\cite{dyr96,dyr06}. On times shorter than structural relaxation, the liquid responds as an elastic solid, and the barrier is assumed to be dominated by the elastic deformation of the surroundings,
\begin{equation}
E(T)\simeq G_\infty(T)V_c,
\end{equation}
where $G_\infty$ is the high-frequency shear modulus and $V_c$ is a characteristic volume. Other approaches relate the barrier to the amplitude of fast cage motion or to spatially resolved elastic properties. Local-elastic descriptions use the response to local force dipoles~\cite{kap21}, while ECNLE constructs a projected free-energy barrier comprising a local cage-escape contribution and a collective elastic cost generated in the surrounding medium~\cite{mir14a,mir14b,pha18}. Relaxation data have been found to be consistent with these different constructions, through correlations with the shear modulus, the Debye--Waller factor, local elastic stiffness, or the cage-plus-elastic barrier~\cite{dyr96,dyr06,lar08,kap21,mei20,zho20,mei21}. At this level, the evidence establishes a robust connection between short-time mechanical properties and slow relaxation, but does not by itself identify which construction captures the origin of the barrier.

The collapse shown in Fig.~\ref{fig:success}(c) provides a particularly broad example. Leporini and co-workers found that relaxation times across diverse glass formers can be organized over many decades by
\begin{equation}
\log_{10}\tau
=
\alpha
+
\beta\left\langle u^2\right\rangle^{-1}
+
\gamma\left\langle u^2\right\rangle^{-2},
\label{eq:generalized_hw}
\end{equation}
where $\langle u^2\rangle$ is the short-time mean-squared displacement, or Debye--Waller factor~\cite{lar08}.
\mpc{They obtained this relation by generalizing the Hall--Wolynes expression, $\tau\propto\exp[a^2/(2\langle u^2\rangle)]$, to a distribution of the squared displacements $a^2$ required for cage escape. The original expression arose within the aperiodic-crystal picture that preceded RFOT~\cite{hal87}, where $\langle u^2\rangle$ measures localization within an amorphous free-energy minimum.}
The same $\langle u^2\rangle$ also admits an elastic interpretation as a measure of cage compliance: in the harmonic regime, $\langle u^2\rangle\sim T/k_{\rm eff}$, with $k_{\rm eff}$ related to the high-frequency stiffness. The Leporini relation itself contains neither the configurational entropy nor a growing thermodynamic length and therefore does not test the thermodynamic content of RFOT. Its compatibility with both aperiodic-localization and elastic interpretations illustrates why even an impressive collapse of relaxation data need not discriminate between different microscopic pictures.

A more direct test of what controls the activation barrier is to identify the nonlinear rearrangements themselves. 
\mpc{In the three-dimensional, swap-equilibrated, continuously polydisperse model with repulsive $r^{-10}$ interactions studied so far, localized excitations measured in inherent structures have a broad energy spectrum $N(E)$ that shifts upward upon cooling~\cite{cia24,jic25}.
In this model, the shift quantitatively tracks the change in activation energy,}
\begin{equation}
E_a(T_1)-E_a(T_2)
\simeq E_g(T_1)-E_g(T_2),
\label{eq:spectrum-shift}
\end{equation}
where $E_g$ characterizes the low-energy scale of $N(E)$. The thermal evolution of the excitation spectrum therefore predicts the change of the activation barrier without an adjustable parameter~\cite{cia24}. This provides a direct microscopic test in which the proposed controlling quantity is measured independently and compared with the barrier itself.

\subsection{Different predictions, different tests}
The theories differ not only in the physical mechanisms they propose, but also in the nature and specificity of their quantitative predictions.
Hierarchical dynamic facilitation commits directly to a specific temperature dependence of the relaxation time: once its parameters have been determined, the parabolic law can be tested by extending measurements of $\tau(T)$ beyond the range over which it was calibrated. 
RFOT and elastic approaches do not, by themselves, prescribe the temperature dependence of $\tau(T)$ without additional material-specific information. 
Instead, they predict quantitative relations between the dynamics and independently measurable quantities: configurational entropy and amorphous static lengths in RFOT, and short-time elastic or vibrational observables in elastic theories. 
Excitation-based approaches place the test at a more microscopic level by asking whether the local barriers can be identified directly and whether their evolution accounts quantitatively for the growth of the macroscopic activation energy. The relevant evidence therefore differs among the theories: the functional form of $\tau(T)$ for hierarchical facilitation, relations between independently measured quantities for thermodynamic and elastic approaches, and a direct correspondence between microscopic barriers and relaxation for excitation-based descriptions.

The role of the microscopic clock $t_0$ also differs among these descriptions. 
RFOT and elastic or Debye--Waller relations contain an extrapolated zero-barrier prefactor whose physical scale can be examined. 
The parabolic law is instead anchored at the measured onset relaxation time $\tau_{\rm o}$. Since $\tau_{\rm o}$ is not a microscopic attempt time, the parabolic representation does not provide a separate prediction for $t_0$; its quantitative commitment concerns the temperature dependence of $\tau$ below the onset.

\section{From successful descriptions to discriminating tests}
\mpc{
Simulations can now equilibrate liquids at lower temperatures and longer relaxation times than were previously accessible, making it possible to test theoretical descriptions beyond the range in which they were established. A central requirement is that a theory should be tested in a way that does not allow disagreement with new data to be absorbed simply through a recalibration of its free or phenomenological parameters. This standard has not always been met: as broader data sets have become available, successful descriptions have often been extended by refitting, which may preserve agreement without providing a sharper test of the underlying mechanism. Two complementary strategies help avoid this ambiguity. One can stress-test an established relation by fixing it in the regime where it was successful and following the equilibrium liquid into a new regime. Alternatively, one can measure independently the microscopic quantities entering a theory and test whether they satisfy the relations the theory requires. The examples below use both strategies.
}

\subsection{Challenges to RFOT}

\paragraph{Dependence on the kinetic rules.}
Swap dynamics provides a direct challenge to thermodynamic descriptions of relaxation: it can accelerate equilibration by many orders of magnitude while preserving the equilibrium measure. Wyart and Cates emphasized that this strongly constrains scenarios in which a growing static length alone fixes the physical relaxation time~\cite{wya17}. 
Within RFOT, Berthier \emph{et al.} proposed the related ``crumbling metastability'' scenario, in which swap postpones metastability and bypasses barriers relevant to the physical dynamics~\cite{ber19b}.

More recently, Ghimenti \emph{et al.} proposed that accelerated algorithms act approximately by reparameterizing time: kinetic rules set the clock while the landscape controls relations among slow observables~\cite{ghimenti2026}. Recent work finds that this reparameterization is only approximate~\cite{PicaCiamarraUnderReview}. The remaining issue is therefore quantitative: how much of the physical dynamics is fixed by equilibrium organization, and how much requires kinetic information?

\paragraph{The exponent $\alpha$.}
A second tension is visible in the generalized Adam--Gibbs relation. The original RFOT construction gives $\alpha=1$, whereas Ref.~\citenum{oza19} favors $\alpha\simeq0.5$--$0.6$ in three dimensions over the experimentally relevant window. Combined with estimates of the mismatch exponent $\theta$, these values imply a barrier exponent $\psi$ smaller than the mean-field value $\psi=\theta=d/2$. Finite-window effects and temperature-dependent prefactors can renormalize effective exponents~\cite{oza19,bir23}, so this is not a falsification of RFOT. It is, however, a numerical value that a microscopic theory should ultimately explain rather than fit.

\paragraph{The microscopic time $t_0$.}
A further quantitative constraint comes from the prefactor. In the activated
form of Eq.~\eqref{eq:tau}, $t_0$ is a microscopic timescale and is
therefore expected to remain tied, at least in order of magnitude, to the
vibrational time $t_{\rm vib}$. Within the generalized RFOT scaling,
Eq.~\eqref{eq:rfot_ag} can be written as
\begin{equation}
\log_{10}\tau=\log_{10}t_0+
\frac{A}{\ln 10}\frac{1}{T s_{\rm c}^{\alpha}},
\label{eq:rfot_intercept}
\end{equation}
so that the intercept of the straight lines in Fig.~\ref{fig:success}(a)
provides an effective estimate of $t_0$. Extrapolating the fitted lines gives
values ranging from roughly $10^{-9}\,$s to $10^{-23}\,$s, with some fits
yielding even smaller values~\cite{VftAnomaliesNote}.
Because this intercept is obtained from a long
extrapolation and is correlated with the fitted exponent $\alpha$, individual
values should not be overinterpreted. Their scale is nevertheless striking:
for $t_0=t_{\rm vib}\exp(-S_a/k_B)$, the smallest values would correspond to
activation entropies of order tens of $k_B$, unless temperature dependences
neglected in the simple scaling form are instead absorbed into the effective
prefactor.
 
A useful independent benchmark comes from a temperature-jump (reheating) protocol that measures the activation barrier and prefactor separately. For one numerical glass former, Ref.~\citenum{cia24} found $t_0\simeq2.4\times10^{-3}t_{\rm vib}$, defining $t_{\rm vib}$ as the time for short-time correlation functions to reach the caging plateau. For comparison, a typical molecular vibrational time is of order $10^{-13}\,$s: an inferred prefactor of $10^{-23}\,$s is therefore about ten orders of magnitude shorter. The measured ratio is model-dependent and need not be universal, but it illustrates the physical scale against which such extrapolated prefactors must be judged. A microscopic theory should account for both the temperature dependence of relaxation and the prefactor it implies.

\paragraph{A direct static test.}
RFOT also permits a stringent test that does not involve dynamics. Equation~\eqref{eq:rfot_xi} relates $s_{\rm c}(T)$, $\xi_{\rm PTS}(T)$, and the mismatch scale $\Upsilon(T)$. All three are now numerically accessible, but to our knowledge they have not yet been determined independently over the same temperature range in a single three-dimensional liquid. Existing work generally constrains only two. For example, Ref.~\citenum{oza19} compared $s_{\rm c}$ and $\xi_{\rm PTS}$ while neglecting a possible temperature dependence of the mismatch scale. A stronger test would measure all three quantities independently and ask whether Eq.~\eqref{eq:rfot_xi} is satisfied, using overlap free-energy methods for the mismatch cost~\cite{Guiselin2022} and established protocols for $\xi_{\rm PTS}$~\cite{BerthierCharbonneauYaida2016}. This would test the static RFOT balance directly, independently of the additional assumptions needed to connect the mosaic length to the relaxation time.

\subsection{Challenges to dynamic facilitation}
\begin{figure}[!t]
\centering
\includegraphics[width=0.45\textwidth]{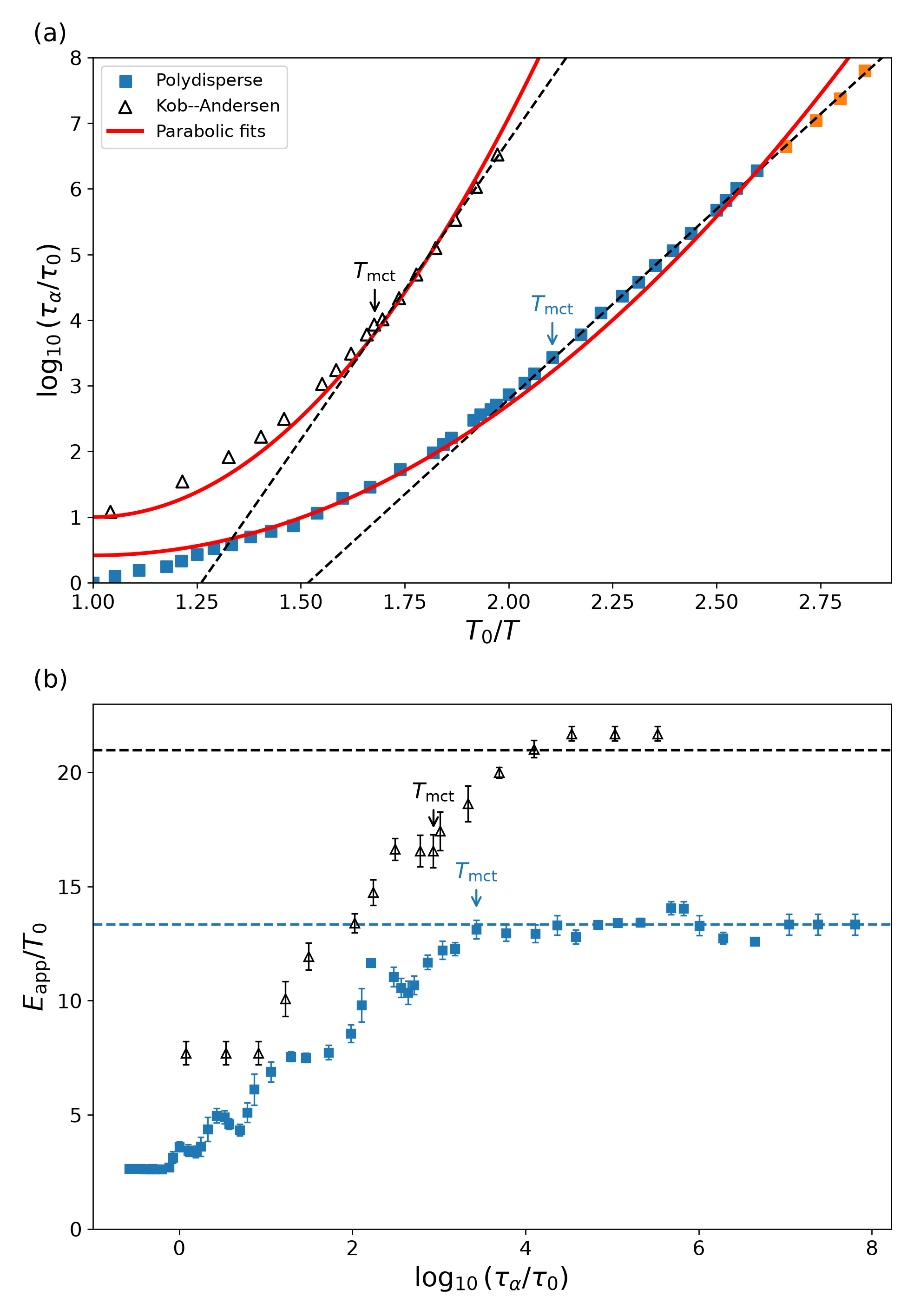}
\caption{(a) Relaxation time versus inverse temperature for the continuously
polydisperse model of Ref.~\citenum{gui22} and the Kob--Andersen mixture of
Ref.~\citenum{Das2022} (shifted vertically for clarity). Orange points for
the polydisperse model are estimates based on time--temperature superposition.
Solid curves show the parabolic prediction, Eq.~\eqref{eq:parabolic}; for the
Kob--Andersen mixture, the parameters are those reported in
Ref.~\citenum{elm10}.
(b) Corresponding apparent activation energy. In both systems, at low
temperature the apparent activation energy ceases to exhibit the increase
upon cooling predicted by the parabolic law and instead approaches a constant,
indicating Arrhenius behavior. Horizontal lines indicate the values obtained
from low-temperature Arrhenius fits in (a).}
\label{fig:parabolic}
\end{figure}
The hierarchical facilitation scenario leads to the parabolic law, Eq.~\eqref{eq:parabolic}. This relation describes transport data for a remarkably broad range of fragile glass formers~\cite{elm09,elm10} and was subsequently connected to localized excitations in atomistic liquids~\cite{keys2011excitations}. It therefore represents a particularly strong and apparently universal prediction of the hierarchical scenario.

Whether the parabolic growth persists at lower temperature has long been debated. Mallamace \emph{et al.} interpreted experimental transport data as evidence for a fragile-to-strong crossover above $T_g$~\cite{Mallamace2010}; Elmatad argued instead that the same data remained consistent with the parabolic law and that the apparent crossover reflected the onset of supercooled dynamics rather than a genuine low-temperature change of regime~\cite{Elmatad2011}, an interpretation subsequently disputed by Mallamace \emph{et al.}~\cite{Mallamace2011Reply}. The published exchange illustrates that the experimentally accessible equilibrium window is too limited to resolve this issue through fits alone.

Deeply equilibrated simulations now allow this issue to be revisited.
Figure~\ref{fig:parabolic}(a) shows two examples suggesting deviations from the parabolic law at low temperature.
For the Kob--Andersen (KA) mixture, the data are from Das and Sastry~\cite{Das2022}. For the continuously polydisperse liquid, the data are from Guiselin~\textit{et al.}~\cite{gui22}, who prepared equilibrium configurations using swap dynamics and subsequently evolved them using the physical dynamics.
The two systems therefore reach the deeply supercooled regime through complementary numerical routes.
Additional examples have been reported for the KA model in Ref.~\citenum{Coslovich2018} and for swap-equilibrated models in Refs.~\citenum{cia24,Parmar2026}.

\mpc{
The deviations in Fig.~\ref{fig:parabolic}(a) also invite a comparison of activation energies. The barrier $E(T)$ entering Eq.~\eqref{eq:tau}, however, cannot be uniquely inferred from relaxation-time data without independent knowledge of the microscopic prefactor $t_0$: a contribution to $E(T)$ proportional to $T$ is indistinguishable from a rescaling of $t_0$. This ambiguity is removed by considering the apparent activation energy,
\begin{equation}
E_{\rm app}(T)=\frac{d\ln\tau}{d(1/T)} = E(T) - T \frac{dE(T)}{dT},
\end{equation}
where the second equality assumes a temperature-independent $t_0$.
The ``Arrhenius'' behavior seen in Fig.~\ref{fig:parabolic}(a) means that $\ln\tau$ is linear in $1/T$, or equivalently that $E_{\rm app}$ is constant.}

The hierarchical facilitation scenario predicts
\begin{equation}
E_{\rm app}(T)
=2J^2\left(\frac{1}{T}-\frac{1}{T_{\rm o}}\right),
\end{equation}
which grows continuously upon cooling. 
By contrast, the finite-difference estimates of $E_{\rm app}$ in Fig.~\ref{fig:parabolic}(b) approach plateaus in both systems: within uncertainty, $E_{\rm app}$ remains constant over about two decades of $\tau$ for the Kob--Andersen mixture and about four decades for the continuously polydisperse liquid. 
These plateaus emerge after an extended super-Arrhenius regime and persist over several decades of relaxation time; they therefore cannot be identified with the onset of supercooled dynamics. This behavior is distinct from the ambiguity that limited the earlier analysis of experimental data.

These simulations, like those that established the parabolic law numerically~\cite{keys2011excitations}, are performed at constant volume, whereas experimental transport data are generally obtained at constant pressure. Even if the crossover does not persist under isobaric conditions or in more fragile liquids, its occurrence in these two systems rules out the hierarchical-facilitation mechanism underlying the parabolic law as a universal explanation for the growth of the activation barrier. Facilitation may still organize how activated events propagate, but in these systems the evidence favors viewing it as a consequence of the underlying activated dynamics rather than its cause.

\subsection{Challenges to elastic and excitation-based descriptions}
\label{sec:elastic-tests}

\paragraph{Discriminating among elastic descriptions.}
Most elastic descriptions rest on ansatzes that relate the activation barrier \mpc{for a finite-amplitude rearrangement that takes the system out of a metastable basin} to a short-time quantity characterizing linear response or fluctuations within that basin: the high-frequency shear modulus in the shoving model~\cite{dyr96,dyr06}, the Debye--Waller factor in Ref.~\citenum{lar08}, and the stiffness associated with the response of the inherent structure to force dipoles~\cite{kap21}.  
These approaches describe $\tau(T)$ comparably well, and the different values of $t_0$ they infer all correspond to physically plausible microscopic clocks. Neither agreement with $\tau(T)$ nor the order of magnitude of $t_0$ therefore discriminates among them. \mpc{Ref.~\citenum{cia24} attempted to go beyond this order-of-magnitude check by developing a method to estimate the microscopic timescale $t_0$, thereby allowing a more stringent comparison between the predicted and measured activation barriers in a model glass former. Once $t_0$ is constrained in this way, the activation energy $E(T)$ can be evaluated and directly compared with the theoretical predictions. This comparison suggests that the barrier inferred from the macroscopic shear modulus increases too weakly upon cooling, whereas that inferred from the average dipolar stiffness increases too strongly~\cite{PicaCiamarra2026local}.}
These quantitative tensions are consistent with the observation that nonlinear local mechanical probes correlate more strongly than linear elastic quantities with the locations of structural rearrangements and, in some cases, with their activation barriers~\cite{li22,ler22}.

These findings shift attention from the average linear response to the local environment in which a rearrangement occurs. 
Locally favoured structures and bond-orientational order grow on cooling and correlate with slow dynamics~\cite{tanaka10,royall15,tong18}. Machine-learning methods likewise distil local structure into a scalar \textit{softness} that predicts rearrangement locations~\cite{cubuk15} and a softness-dependent activation scale in glassy liquids~\cite{schoenholz16}. Such correlations suggest that local structure may help determine where rearrangements occur and how costly they are, although they do not by themselves specify the barrier.

ECNLE attempts to construct that barrier from the equilibrium structure, combining a nonlinear local cage-escape contribution with a collective elastic cost~\cite{mir14a,mir14b,mirigian15,pha18}. For hard spheres these contributions are calculated from the equilibrium structure, but applications to molecular liquids rely on an empirical thermodynamic mapping to an effective hard-sphere fluid.

\paragraph{Measuring nonlinear barriers directly.}
This motivates identifying and measuring the nonlinear rearrangements directly.
Localized excitations can now be resolved in inherent structures, where their energies
form a broad spectrum $N(E)$ whose low-energy scale shifts upward on cooling; in the
model studied so far this shift quantitatively tracks the growth of the activation
energy, Eq.~\eqref{eq:spectrum-shift}, with no adjustable parameter~\cite{cia24,jic25}.
The correspondence has been established essentially for one model; it predicts the temperature dependence of the barrier and not its absolute value; and inherent-structure methods access
potential-energy barriers rather than finite-temperature free-energy barriers.
Establishing transferability across glass formers, constructing finite-temperature
excitation spectra, and connecting elementary barriers to irreversible structural
relaxation are therefore the natural next steps.

\section{Outlook: what could be done}
\label{sec:outlook}
Progress in understanding the glassy slowdown requires quantitative tests of proposed mechanisms. The quantities entering a theory should be measured independently and used to test its predictions. When possible, this should be complemented by interventions that modify a proposed controlling mechanism while leaving other relevant properties unchanged. The first type of test establishes whether a theoretical relation is satisfied; the second can determine whether the quantity it identifies plays a causal role in the slowdown.

\paragraph{Close the RFOT relations.}
The RFOT construction makes it possible to test its static and dynamical assumptions separately. Equation~\eqref{eq:rfot_xi} relates $s_{\rm c}(T)$, $\xi_{\rm PTS}(T)$, the mismatch scale $\Upsilon(T)$, and the exponent $\theta$ governing the size dependence of the mismatch cost. Configurational entropy, point-to-set lengths, and mismatch free energies are now numerically accessible~\cite{BerthierCharbonneauYaida2016,Guiselin2022}. A complete static test, however, also requires determining $\theta$ independently, rather than assuming its value or inferring it from the relaxation dynamics. Measuring these quantities over the same temperature and length ranges in a single three-dimensional liquid would establish whether Eq.~\eqref{eq:rfot_xi} is satisfied.

If the static relation holds, one could then test separately whether the measured mosaic length controls the activation barrier according to Eq.~\eqref{eq:rfot_barrier}. This requires determining the barrier exponent $\psi$ and the microscopic barrier scale independently. The measured values of $\theta$ and $\psi$ would then predict
$\alpha=\psi/(d-\theta)$, allowing the fitted value $\alpha\simeq0.5$--$0.6$ to be confronted with RFOT rather than treated as an adjustable exponent. Such a program would distinguish a failure of the static mosaic balance from a failure of the dynamical assumption connecting the mosaic length to relaxation. It should also account for the physical scale of the prefactor, rather than leaving extrapolated times as small as $10^{-23}$s unexplained.

\paragraph{Test the low-temperature crossover under different conditions.}
An important next step is to test whether the approach to Arrhenius behavior persists along isobaric paths and in more fragile liquids closer to experimental systems. Isobaric cooling often produces a more fragile temperature dependence, so that the crossover may occur at temperatures that are more difficult to reach~\cite{sim26}. If it persists, the hierarchical mechanism underlying the parabolic law cannot provide a universal explanation of the barrier even under these conditions. If it does not, the present observations would instead reveal a dependence on thermodynamic path or on the class of glass former, helping to explain the success of the parabolic law for experimental data.

\paragraph{Quantify facilitation and test whether it is rate-limiting.}
The validity of the parabolic law and the presence of facilitation are distinct questions. Facilitation is a causal concept: a relaxation event facilitates another if its occurrence increases the subsequent relaxation rate nearby beyond that expected from the initial configuration. 
Existing evidence, however, is based largely on spatiotemporal correlations. 
Short-time cage jumps form clusters and avalanches that build large-scale dynamical heterogeneity~\cite{candelier09granular,candelier10liquid}, while mobility propagates preferentially through soft regions identified by isoconfigurational maps~\cite{candelier10liquid}. 
These observations establish correlated propagation, but do not distinguish causal triggering from repeated activity within regions that were already more likely to relax. 
Despite the central role assigned to facilitation, there is no generally accepted quantitative measure that separates these contributions. 
A natural target would be the fraction of relaxation events attributable to preceding nearby events, evaluated relative to the event rate expected from the initial configuration had those events not occurred. 
Constructing such a measure remains an open challenge. 
Without it, one cannot establish whether facilitation strengthens upon cooling, compare its importance across glass formers, or test whether it tracks fragility or the growth of the activation barrier.

Even if such a measure revealed a nonzero fraction of causally triggered events, this would not establish that facilitation controls $\tau$. The rate-limiting step could instead be the creation of localized events, with facilitation organizing their subsequent propagation. A decisive test would modify the propagation of mobility while preserving the emergent stationary distribution of local barriers. If $\tau$, and particularly its leading activation barrier, changed under such an intervention, this would demonstrate that facilitation contributes to setting the relaxation rate. If only dynamical correlations and heterogeneity changed, facilitation would instead organize relaxation without being rate-limiting.
Thermal elastoplastic models offer a natural setting for such a test, since the activation rule and elastic propagator are explicit\cite{oza23,tah23}. Extending these models to vary the propagation of mobility while preserving the emergent barrier distribution would isolate the effect of propagation on the relaxation time.

\paragraph{Extend the excitation-spectrum test.}
The correspondence between the excitation spectrum and the activation barrier, Eq.~\eqref{eq:spectrum-shift}, provides a direct test of a proposed microscopic origin of the slowdown. Its range of validity, however, is not yet known. The correspondence has been established essentially for one model, predicts the change of the barrier rather than its absolute value, and concerns potential-energy barriers in inherent structures rather than finite-temperature free-energy barriers. Constructing $N(E)$ for glass formers with different interaction softness, bonding, and fragility would establish whether the same relation is transferable or applies only to a restricted class of systems. 
This program is feasible in principle, but numerically very intensive.
Finite-temperature excitation spectra are also needed to determine whether the correspondence survives when entropic contributions become relevant. 

\mpc{
\paragraph{Experimental routes to discriminating tests.}
This Perspective has focused primarily on numerical studies, because simulations provide direct access to microscopic quantities and allow controlled tests that are difficult to realize experimentally. 
However, simulations necessarily rely on idealized model glass formers, and whether the mechanisms identified numerically actually operate in real glass-forming materials must ultimately be established experimentally.}

\mpc{There is also a practical reason why experiments remain crucial. Recent algorithmic advances have largely closed the gap with experiments in the ability to equilibrate deeply supercooled states, at least for suitably chosen polydisperse model systems. They have not, however, closed the corresponding gap in the dynamics: once these deeply equilibrated configurations are evolved with the physical dynamics, relaxation becomes prohibitively slow at the lowest accessible temperatures. Experiments can therefore probe dynamical regimes that remain difficult to access numerically.
In this spirit, I highlight below two experimental directions that could help address some of the open questions discussed throughout this Perspective.}

\mpc{
The first concerns the interplay between the activation barrier $E(T)$ and the microscopic timescale $t_0$, a recurring issue in the preceding discussion. Measurements of the equilibrium relaxation time determine only the combination $\tau(T)=t_0\exp[E(T)/T]$, so that the barrier cannot be reconstructed uniquely without independent information on $t_0$. Ref.~\citenum{cia24} addressed this problem numerically by means of a temperature-jump protocol. A configuration equilibrated at a preparation temperature $T_0$ was evolved after an abrupt change to a different temperature $T_1$. Over a regime in which the structure still reflects the preparation temperature, the relaxation time is expected to obey $\tau(T_1|T_0)=t_0\exp[E(T_0)/T_1]$. Repeating the protocol for several values of $T_1$ therefore separates $E(T_0)$, obtained from the slope of $\ln\tau$ versus $1/T_1$, from the microscopic prefactor $t_0$. An experimental implementation of this protocol in molecular glass formers would provide a direct determination of the microscopic clock and of the temperature dependence of the activation barrier. Such measurements would remove an ambiguity that presently limits comparisons among thermodynamic, elastic, and excitation-based descriptions, for which similar relaxation-time data may correspond to different decompositions between barrier and prefactor.}

\mpc{
The second possibility concerns two-dimensional or quasi-two-dimensional colloidal glass formers, where individual particles can be imaged and manipulated. 
Holographic optical tweezers can generate independently addressable arrays of traps, potentially allowing the two species of a bidisperse colloidal glass former to be identified, positioned at prescribed coordinates, and confined with separately calibrated trap strengths~\cite{curtis02,chapin06}; related techniques have already been used to impose amorphous pinned boundaries in two-dimensional colloidal glass formers~\cite{nagamanasa15}. One could therefore envisage using spatially patterned optical trapping to prepare repeatedly essentially the same particle configuration and then release it to generate multiple dynamical realizations from the same initial structure, providing an experimental analogue of an isoconfigurational ensemble. Such a protocol would separate structural propensity from dynamical triggering and thereby test facilitation more directly: one could compare the probability of a rearrangement following a nearby event with the corresponding probability in realizations in which that event did not occur. Importantly, these measurements could be performed at degrees of overcompression for which following the physical relaxation dynamics is already prohibitively expensive in atomistic simulations. They could therefore probe how facilitation evolves deep in the glassy regime, where numerical evidence is presently most limited, and move its experimental characterization from spatiotemporal correlation toward a genuinely causal test.}
\\
\\
Much of the search for a theory of the glass transition implicitly assumes that the same mechanism underlies the slowdown of different glass formers. This assumption is not guaranteed. Broadly similar glassy phenomenology can arise in models with essentially trivial thermodynamics as well as in systems displaying nontrivial structural and thermodynamic correlations. The similarity of the resulting dynamics therefore does not imply a common microscopic origin. Indeed, signatures associated with dynamic facilitation and thermodynamic scenarios can coexist in the same deeply supercooled system~\cite{ortlieb23}, but their coexistence does not establish that a common mechanism controls the barrier. Different mechanisms may dominate in different liquids, or even in different dynamical regimes of the same liquid. A useful theory should consequently establish not only whether a proposed mechanism can reproduce the slowdown, but also where it does so quantitatively and where it fails. The field should remain open to the possibility that there is no single universal mechanism of glass formation.

\begin{acknowledgments}
I acknowledge discussions with colleagues in the glass and soft-matter communities. 
This work has been supported by the Singapore Ministry of Education through grants
MOE-T2EP50224-0016, RG152/23, and RG153/24. 
\end{acknowledgments}

\section*{Author Declarations}

\subsection*{Conflict of Interest}
The author has no conflicts to disclose.

\subsection*{Data Availability}
The data underlying the figures are available from the cited publications. 

\bibliography{perspective_references}

\end{document}